\documentclass[sigconf, nonacm]{acmart}

\newcommand\vldbdoi{XX.XX/XXX.XX}
\newcommand\vldbpages{XXX-XXX}
\newcommand\vldbvolume{14}
\newcommand\vldbissue{1}
\newcommand\vldbyear{2020}
\newcommand\vldbauthors{\authors}
\newcommand\vldbtitle{\shorttitle}

\usepackage{graphicx}
\usepackage{balance}  
\usepackage{xcolor}
\usepackage{xspace}
\usepackage{multirow}
\usepackage{url}
\usepackage{newfloat}
\usepackage[noabbrev]{cleveref}
\usepackage[labelfont=bf]{caption}
\usepackage[labelfont=bf]{subcaption}
\usepackage{enumitem}
\usepackage{textcomp}
\usepackage{etoolbox}
\usepackage{listings}
\usepackage[subtle]{savetrees}
\usepackage{algorithm}
\usepackage{algorithmic}

\newtoggle{rcolors} \togglefalse{rcolors}
\newcommand{\colora}[1]{\iftoggle{rcolors}{{\color{red}{#1}}}{#1}}

\newcommand{\sn}{\textsc{Smol}\xspace}
\newcommand{\noscope}{\textsc{NoScope}\xspace}
\newcommand{\blazeit}{\textsc{BlazeIt}\xspace}
\newcommand{\tahoma}{\textsc{Tahoma}\xspace}

\newcommand{\minihead}[1]{{\vspace{.4em}\noindent\textbf{#1.} }}
\newcommand{\miniheadit}[1]{{\vspace{.4em}\noindent\textit{#1.} }}

\begin{document}


\title{Jointly Optimizing Preprocessing and Inference for DNN-based Visual Analytics}

\author{
Daniel Kang, Ankit Mathur, Teja Veeramacheneni, Peter Bailis, Matei Zaharia
}
\affiliation{
  \institution{Stanford DAWN Project}
}

\begin{abstract}

While deep neural networks (DNNs) are an increasingly popular way to query large
corpora of data, their significant runtime remains an active area of research.
As a result, researchers have proposed systems and optimizations to reduce these
costs by allowing users to trade off accuracy and speed.
In this work, we examine \emph{end-to-end} DNN execution in visual
analytics systems on modern accelerators. Through a novel measurement
study, we show that the \emph{preprocessing of data} (e.g., decoding, resizing)
can be the bottleneck in many visual analytics systems on modern hardware.

To address the bottleneck of preprocessing, we introduce two optimizations for
\emph{end-to-end} visual analytics systems. First, we introduce novel methods of
achieving accuracy and throughput trade-offs by using natively present,
low-resolution visual data. Second, we develop a runtime engine for efficient
visual DNN inference.  This runtime engine a) efficiently pipelines
preprocessing and DNN execution for inference, b) places preprocessing
operations on the CPU or GPU in a hardware- and input-aware manner, and c)
efficiently manages memory and threading for high throughput execution. We
implement these optimizations in a novel system, \sn, and evaluate \sn on eight
visual datasets. We show that its optimizations can achieve up to 5.9$\times$
\emph{end-to-end} throughput improvements at a fixed accuracy over recent work
in visual analytics.

\end{abstract}

\maketitle

\begingroup\small\noindent\raggedright\textbf{PVLDB Reference Format:}\\
\vldbauthors. \vldbtitle. PVLDB, \vldbvolume(\vldbissue): \vldbpages, \vldbyear.\\
\href{https://doi.org/\vldbdoi}{doi:\vldbdoi}
\endgroup
\begingroup
\renewcommand\thefootnote{}\footnote{\noindent
This work is licensed under the Creative Commons BY-NC-ND 4.0 International License. Visit \url{https://creativecommons.org/licenses/by-nc-nd/4.0/} to view a copy of this license. For any use beyond those covered by this license, obtain permission by emailing \href{mailto:info@vldb.org}{info@vldb.org}. Copyright is held by the owner/author(s). Publication rights licensed to the VLDB Endowment. \\
\raggedright Proceedings of the VLDB Endowment, Vol. \vldbvolume, No. \vldbissue\ %
ISSN 2150-8097. \\
\href{https://doi.org/\vldbdoi}{doi:\vldbdoi} \\
}\addtocounter{footnote}{-1}\endgroup

\section{Introduction}

Deep neural networks (NNs) now power a range of visual analytics tasks and
systems~\cite{kang2017noscope, hsieh2018focus, anderson2018predicate,
kang2019blazeit} due to their high accuracy, but state-of-the-art DNNs can be
computationally expensive. For example, accurate object detection methods can
execute as slow as 3-5 frames per second (fps)~\cite{he2017mask,
tan2019efficientdet}.

\sloppypar{
To execute visual analytics queries efficiently, systems builders have developed
optimizations to trade off accuracy and throughput \cite{kang2017noscope,
anderson2018predicate, hsieh2018focus, kang2019blazeit, lu2018accelerating}:
more accurate DNNs are more computationally expensive~\cite{he2016deep,
tan2019efficientnet, tan2019efficientdet}. Many of these systems (e.g.,
\noscope, \blazeit, \tahoma, and probablistic predicates) accelerate visual
analytics queries by using proxy or \emph{specialized NNs}, which approximate
larger target DNNs. These specialized NNs can be up to 5 orders of magnitude
cheaper to execute than their target DNNs and are used to filter inputs so the
target DNNs will be executed fewer times~\cite{kang2017noscope,
anderson2018predicate, hsieh2018focus, kang2019blazeit, lu2018accelerating}.
}


This prior work focuses solely on reducing DNN execution time. These systems
were built before recent DNN accelerators were introduced and were thus
benchmarked on older accelerators. In this context, these systems correctly
assume that DNN execution time is the overwhelming bottleneck. For example,
\tahoma benchmarks on the NVIDIA K80 GPU, which executes ResNet-50 (a
historically expensive DNN~\cite{mlperf2018, coleman2017dawnbench,
coleman2018analysis}) at 159 images/second.

However, as accelerators and compilers have advanced, these systems ignore a key
bottleneck in \emph{end-to-end} DNN inference: preprocessing, or the process of
decoding, transforming, and transferring image data to accelerators. In the
first measurement study of its kind, we show that preprocessing costs often
\emph{dominate end-to-end DNN inference} when using advances in hardware
accelerators and compilers. For example, the historically expensive
ResNet-50~\cite{coleman2017dawnbench, mlperf2018} has improved in throughput by
28$\times$ on the inference-optimized NVIDIA T4 GPU. As a result, ResNet-50 is
now 9$\times$ higher throughput than CPU-based image preprocessing, making
\emph{preprocessing the bottleneck}, on the inference-optimized
\texttt{g4dn.xlarge} Amazon Web Services (AWS) instance, which has a NVIDIA T4.
This boost in efficiency translates to both power and dollar costs:
preprocessing requires approximately 2.3$\times$ as much power and costs
11$\times$ as much as DNN execution (\S\ref{apx:hardware}). Similar
results hold for Google Cloud's T4 inference optimized instances. These
imbalances become only higher with smaller specialized NNs that recent visual
analytics systems use.


In light of these observations, we examine opportunities for more principled
\emph{joint optimization} of preprocessing and DNN execution. We leverage two
insights: a) the accuracy and throughput of a DNN is closely coupled with its
input format and b) preprocessing operations can be placed on both CPUs and
accelerators. Thus, rather than treating the input format as fixed, we consider
methods of using inputs as a key step in DNN architecture search and training.

This yields two novel opportunities for accelerating inference: a) cost-based
methods that leverage low-resolution visual data for higher accuracy or improved
throughput and b) input- and hardware-aware methods of placing preprocessing
operations on the CPU or accelerator and correctly pipelining computation.

A critical component to leverage these opportunities is a cost model to select
query plans. We correct the erroneous assumption in prior work that DNN
execution dominates \emph{end-to-end} DNN inference. We instead propose a
cost model that is preprocessing aware and validate that our cost model is more
accurate than prior cost models. While our preprocessing aware cost model is
simple, it enables downstream optimizations, described below.

First, we propose methods of using natively present, low resolution visual data
for more efficient, input-aware accuracy/throughput trade offs. Image and video
serving sites often have natively present low resolution data, e.g., Instagram
has thumbnails~\cite{arens2019always} and YouTube stores multiple resolutions of
the same video. Even when low resolution data is not natively present, we can
partially decode visual data (e.g., omitting the deblocking filter in H.264
decoding). As such, we can use natively present data or
partial decoding for reduced preprocessing costs. However, naively using this
reduced fidelity data can reduce accuracy. To recover accuracy, we propose an
augmented DNN training procedure that explicitly uses data augmentation for the
target resolution. Furthermore, we show that using larger, more accurate DNNs
on low resolution data can result in higher accuracy than smaller DNNs on full
resolution data. Enabled by our new preprocessing-aware cost model, we can select
input formats and DNN combinations that achieve better accuracy/throughput trade
offs.


Second, we decide to place preprocessing operations on the CPU or accelerator
to balance the throughput of DNN execution and preprocessing. Furthermore, to
enable high-performance pipelined execution, we build an optimized runtime
engine for end-to-end visual DNN inference. Our optimized runtime engine makes
careful use of pipelined execution, memory management, and high-performance
threading to fully utilize available hardware resources.

We implement these optimizations in \sn, a runtime engine for end-to-end DNN
inference that can be integrated into existing visual analytics systems. We use
\sn to implement the query processing methods of two modern visual analytics
systems, \blazeit~\cite{kang2019blazeit} and
\tahoma~\cite{anderson2018predicate}, and evaluate \sn on eight visual datasets,
including video and image datasets. We verify our cost modeling choices
through benchmarks on the public cloud and show that \sn can achieve up to
5.9$\times$ improved throughput on recent GPU hardware compared to recent work
in visual analytics.

In summary, we make the following contributions:
\begin{enumerate}[itemsep=0em,parsep=0em,topsep=0em, leftmargin=1.5em]
  \item We show that preprocessing costs can dominate end-to-end DNN-based
  analytics when carefully using modern hardware.

  \item We illustrate how to use natively-encoded low-resolution visual data
  formats and specialized NNs to achieve input-aware accuracy/throughput
  trade-offs.

  \item We propose and implement methods of further balancing preprocessing and
  DNN execution, including hardware- and input-aware placement of preprocessing
  operations.

\end{enumerate}

\section{Measurement~Study~of~End-to-End\\DNN~Inference}
\label{sec:benchmarks}

We benchmark DNNs and visual data preprocessing on the public cloud. Our
results show that new accelerators have dramatically improved throughputs and
reduced both dollar and power costs of DNN execution. Given these reduced costs,
we show that \emph{preprocessing costs} can now dominate when using these
accelerators efficiently, even for DNNs that have been considered expensive.

In this section, we benchmark throughputs on the inference-optimized T4 GPU with
a dollar cost-balanced number of vCPU cores on an AWS instance. Our benchmarks
show that preprocessing dominates in both dollar cost and power costs. For
example, preprocessing requires $2.2\times$ as much power (158W vs 70W) and
costs $11\times$ as much for ResNet-50 (\$2.37 vs \$0.218,
\S\ref{apx:hardware}). These trends are similar for other cloud providers (e.g.,
Google Cloud Platform's T4-attachable instances and Microsoft Azure's newly
announce T4 instances) and instance types.

\minihead{Experimental setup}
We benchmarked the popular ResNet-50 model for image
classification~\cite{he2016deep}, which has widely been used in
benchmarking~\cite{coleman2017dawnbench, mlperf2018} and has been considered
expensive. Specialized NNs are typically much cheaper than ResNet-50.

We benchmarked the time for only DNN execution and the time for preprocessing
separately to isolate bottlenecks.

We benchmarked on the publicly available inference-optimized NVIDIA T4
GPU~\cite{nvidia2020t4}. We used the \texttt{g4dn.xlarge} AWS instance which has
4 vCPU cores (hyperthreads): this configuration is cost balanced between vCPUs
and the accelerator (\S\ref{apx:hardware}). This instance type is
optimized for DNN inference; similar instances are available on other cloud
providers. We used the TensorRT compiler~\cite{nvidia2019tensort} for optimized
execution. While we benchmarked on the T4, other contemporary, non-public
accelerators report similar or improved results~\cite{fowers2018configurable,
jouppi2017datacenter}.

\minihead{Effect of software on throughput}
We benchmarked ResNet-50 throughput on the inference-optimized T4 GPU using
three software systems for DNNs to show how more efficient software affects throughput. We benchmark using Keras~\cite{chollet2015keras},
PyTorch~\cite{paszke2017automatic}, and TensorRT~\cite{nvidia2019tensort}. We
note that Keras was used by \tahoma and TensorRT is an optimized DNN
computational graph compiler.

As shown in Table~\ref{table:framework-throughput}, efficient use of
accelerators via optimized compilers (TensorRT) can result
in up to a 10$\times$ improvement in throughput. Importantly, preprocessing
becomes the bottleneck with the efficient use of accelerators.

\begin{table}
\centering
\setlength\itemsep{2em}
\small
\begin{tabular}{ll}
Execution environment & Throughput (im/s) \\ \hline
Keras     & 243 \\
PyTorch   & 424 \\
TensorRT  & 4,513
\end{tabular}
\caption{Throughput of ResNet-50 on the T4 with three different execution
environments. Keras was used in~\cite{anderson2018predicate}. The
efficient use of hardware can result in over a 17$\times$ improvement in
throughput. We used the optimal batch size for each framework (64, 256, and 64
respectively).}
\label{table:framework-throughput}
\vspace{-1.5em}
\end{table}

\minihead{Breakdown of end-to-end DNN inference}
DNN inference consists of steps beyond the execution of the DNN computation
graph. For the standard ResNet-50 configuration, the preprocessing steps
are~\cite{he2016deep, mattson2019mlperf}:
\begin{enumerate}[itemsep=0em,parsep=0em,topsep=0em]
  \item Decode the compressed image, e.g., JPEG compressed.

  \item Resize the image with an aspect-preserving resize such that the short
  edge of the image is 256 pixels. Centrally crop the image to 224x224.

  \item Convert the image to float32. Divide the pixel values by 255, subtract a
  per-channel value, and divide by a per-channel value (these values are derived
  from the training set), i.e., ``normalizing'' the image.

  \item Rearrange the pixel values to channels-first (this step depends
  on the DNN configuration).
\end{enumerate}

To see the breakdown of preprocessing the costs, we implemented these
preprocessing steps in hand-optimized C++, ensuring best practices for high
performance C++, including reuse of memory to avoid allocations. We used
\texttt{libturbo-jpeg}, a highly optimized library for JPEG decompression, for
decoding the JPEG images. We used OpenCV's optimized image processing libraries
for the resize and normalization. For DNN execution, we executed the DNNs with
TensorRT and multiple CUDA streams on synthetic images. We run all benchmarks on
a standard \texttt{g4dn.xlarge} AWS instance and use multithreading to utilize
all the cores.

\begin{figure}[t!]
  \includegraphics[width=\columnwidth]{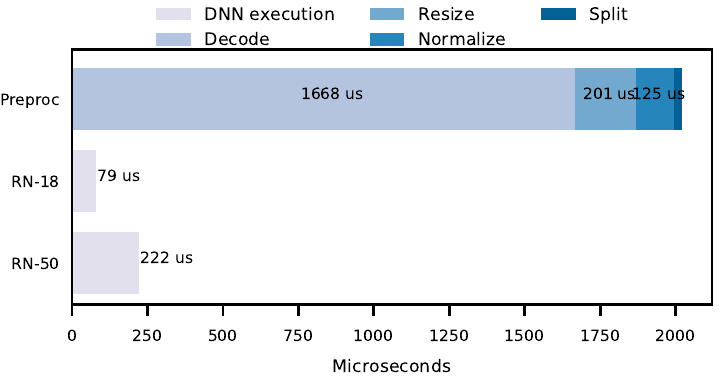}
  \vspace{-0.5em}
  \caption{Breakdown of end-to-end inference per image of ResNet-50 and
  18 for a batch size of 64 on the inference-optimized AWS \texttt{g4dn.xlarge}
  instance type (one NVIDIA T4 GPU and 4 vCPU cores). The DNN was executed on
  the T4 and the preprocessing was parallelized across all CPU cores. The
  execution of the DNN is 7.1$\times$ and 22.9$\times$ faster than preprocessing
  data for ResNet-50 and 18 respectively.}
  \label{fig:bs64-naive}
  \vspace{-1.5em}
\end{figure}

As shown in Figure~\ref{fig:bs64-naive}, simply decoding the JPEG files achieves lower
throughput than the throughput of ResNet-50 execution. All together, the
preprocessing of the data achieves 7.1$\times$ lower throughput than ResNet-50
execution. These overheads increase to up to 22.9$\times$ for ResNet-18.
As discussed, preprocessing dominates in terms of power and dollar costs as
well.

Similar results hold for other networks, such as the
MobileNet-SSD~\cite{liu2016ssd, howard2017mobilenets} used by MLPerf
Inference~\cite{reddi2019mlperf}. This DNN executes at 7,431 im/s, compared to a
preprocessing throughput of 397 im/s on the MS-COCO dataset.

\minihead{Discussion}
Several state-of-the-art DNNs execute far slower than the DNNs benchmarked in
this section, e.g., a large Mask R-CNN may execute at 3-5 fps. However, many
systems use specialized NNs to reduce invocations of these large DNNs. For
example, the \blazeit system uses a specialized NN to approximate the larger
DNN, which reduces the number of large DNN invocations~\cite{kang2019blazeit}.
As these specialized NNs are small (potentially much smaller than even
ResNet-50), we believe our benchmarks are of wide applicability to DNN-based
visual analytics.


\section{\sn Overview}
\label{sec:sys-overview}

\begin{figure}[t!]
  \includegraphics[width=\columnwidth]{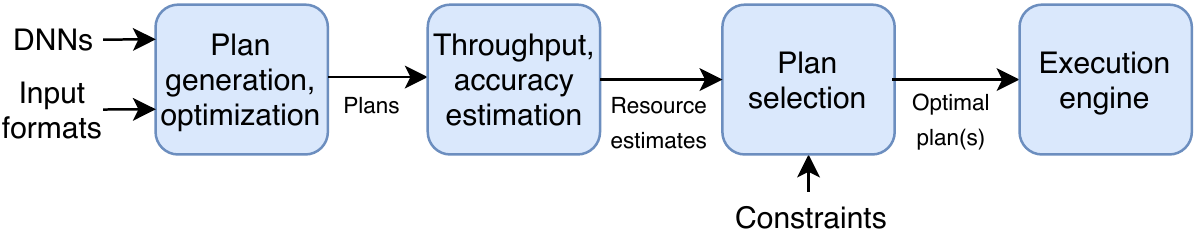}
  \vspace{-0.5em}
  \caption{System diagram of \sn. As input, \sn takes a set of DNNs, visual
  input formats, and optional constraints. As output, \sn returns an optimal set
  of plans or plan, depending on the constraints. \sn will generate plans,
  estimate the resources for each plan, and select the Pareto optimal set of
  plans.}
  \label{fig:sys-diagram}
  \vspace{-1.5em}
\end{figure}

To reduce the imbalance between preprocessing and DNN execution, we develop a
novel system, \sn. \sn's goal is to execute \emph{end-to-end} batch visual
analytics queries. Unlike prior work, \sn aims to optimize end-to-end query
time, including the computational cost of preprocessing in addition to the
computational cost of DNN execution.

To execute these visual analytics queries, \sn uses a cost-based model to
generate query plans that span preprocessing and DNN execution. \sn executes
these plans in its optimized end-to-end inference engine. For a given query
system (e.g., \tahoma or \blazeit), \sn's cost model must be integrated into the
system.

We show a schematic of \sn's architecture in Figure~\ref{fig:sys-diagram}.

\subsection{System Overview}

\minihead{Deployment setting}
In this work, we focus on high throughput batch settings, as recent work
does~\cite{anderson2018predicate, kang2019blazeit}. \sn's goal is to achieve the
highest throughput on the available hardware resources. For example, a visual
analytics engine might ingest images or videos daily and run a batch analytics
job each night. \colora{\sn is most helpful for preprocessing-bound workloads on
large datasets. As we describe, \sn accepts models exported from training
frameworks (e.g., PyTorch, TensorFlow, or Keras) and optimizes its inference. As
such, it is designed to be used at inference time, not with training
frameworks.}

Nonetheless, several of our techniques, particularly in jointly
optimizing preprocessing and inference, also apply to the low-latency or
latency-constrained throughput settings.

We note that in high throughput batch settings, visual data is almost always
stored in compressed formats that require preprocessing. Uncompressed visual
data is large: a single hour of 720p video is almost 900GB of data whereas
compressed video can be as small as 0.5GB per hour. Similarly, JPEG images can
be up to 10$\times$ smaller than uncompressed still images.

\minihead{\sn inference}
As inputs, \sn will take a set of trained DNNs and a set of natively available
visual data formats (e.g., full resolution JPEG images, thumbnail JPEGs). We
denote the set of DNNs as $\mathcal{D}$ and the set of visual data formats as
$\mathcal{F}$. \sn further takes a set of calibration images (i.e., a validation
set) to estimate accuracy.

Given these inputs, \sn will estimate costs to select a plan (concretely, a DNN
and an input format). \sn will then optimize this plan and execute it.

\sn optionally takes a throughput or accuracy constraint at inference time. If a
constraint is specified, \sn will select an optimized execution plan that
respects these constraints. Otherwise, \sn will execute the highest throughput
plan. \sn can be integrated with other systems by returning a Pareto optimal set
of plans (in accuracy and throughput). The calling system will then select a
plan that \sn will execute.

\minihead{\sn training}
While the user can provide the set of trained DNNs, \sn can optionally train
specialized NNs as well. Given a set of DNN architectures (e.g., ResNets) and
the natively available formats, \sn will choose to train some or all of the
DNNs. Given the initial set of models on full resolution data, \sn will
fine-tune the networks on the cross product of $\mathcal{D}$ and resolutions
(\sn will use the same model for different formats of the same resolution).
As \sn fine-tunes, this process adds at most a 30\% overhead in training in
the settings we consider. \sn can also train these network at execution
time~\cite{kang2019blazeit}.

\minihead{Components}
\sn implements the training phase as other systems do~\cite{kang2019blazeit,
kang2017noscope, anderson2018predicate}. As training specialized NNs has been
studied in depth in prior work, we defer discussion to this prior work. \sn
differs from these systems only in it's low-resolution augmented training
(discussed below).

At inference time, \sn contains three major components: 1) a plan generator, 2) a
cost estimator, and 3) an execution engine. We show these components in
Figure~\ref{fig:sys-diagram}.

\sn first generates query plans from $\mathcal{D}$ and $\mathcal{F}$ by taking
$\mathcal{D} \times \mathcal{F}$. For each plan, \sn will estimate the relative
costs of preprocessing and DNN execution and decide where to place preprocessing
operations (i.e., on the CPU or accelerator) for highest throughput. Given these
optimized plans, \sn will estimate the accuracy and throughput of these plans
using its cost model. This process is cheap compared to training, so \sn
exhaustively benchmarks the Pareto frontier of $\mathcal{D} \times
\mathcal{F}$. \sn uses a preprocessing-aware cost model, in contrast to
prior work that ignores these costs. Finally, \sn will return the best query
plan if a constraint is specified or the Pareto optimal set of query plans if
not.

\minihead{Optimizations}
To efficiently execute queries, \sn has several optimizations for improved
accuracy/throughput trade offs and an efficient DNN execution engine.

Briefly, \sn achieves improved accuracy and throughput trade offs by considering
an expanded set of DNNs and leveraging natively present low-resolution data
(\S\ref{sec:acc-thpt}). In contrast, prior work considers only one input
format. From the selected DNN and input format, \sn will efficiently execute
such plans by placing preprocessing operations on CPUs or accelerators in a
hardware- and input-aware manner, efficiently pipelining computation stages, and
optimizing common preprocessing operations (\S\ref{sec:preproc-opt}). We
describe these optimizations in detail below.

\subsection{Examples}
\minihead{Classification example}
\sn can be incorporated into prior work that uses specialization for
classification queries~\cite{kang2017noscope, anderson2018predicate,
lu2018accelerating}. These queries are often binary classification queries,
e.g., the presence or absence of a car in a video. We describe \tahoma in this
example, but note that other systems are similar in spirit.

\tahoma uses a fixed target model and considers a fixed input format, namely the
provided input format of full-resolution JPEG images. \tahoma considers 24
specialized NN, each of which are cascaded with the target DNN. Thus,
$|\mathcal{F}| = 1$ and $|\mathcal{D}| = 24$. \tahoma aims to return the
configuration with the highest throughput for a given accuracy. \tahoma
estimates the throughput of $D_i \in \mathcal{D}$ by \emph{adding} preprocessing
costs, which we show leads to inaccurate throughput estimates. We further note
that \tahoma considers downsampling full resolution images for improved DNN
execution, but not for reduced preprocessing costs.

In contrast, \sn can use natively present thumbnail images, which would expand
$\mathcal{F}$. Decoding these thumbnail images is significantly more efficient,
resulting in higher throughput.

\minihead{Aggregation example}
\sn can be incorporated into prior work that uses specialized
NNs for aggregation queries over visual data, e.g., the number of cars in a
video. The recent \blazeit system uses specialized NNs as a control variate to
reduce the variance in sampling~\cite{kang2019blazeit}. As the variance is
reduced, this procedure results in fewer target model invocations compared to
standard random sampling. \blazeit trains a single specialized NN
($|\mathcal{D}| = 1$) and uses a fixed input format ($|\mathcal{F}| = 1$).

In contrast, \sn can use an expanded set of videos which are encoded at
different resolutions. Namely, \sn considers $|\mathcal{F}| > 1$. These other
formats are natively present in many serving applications, e.g., for thumbnail
or reduced bandwidth purposes.

\vspace{0.3em}

We describe \sn's cost model, optimizations, its implementation, and its
evaluation below.

\section{Cost Modeling for Visual Analytics}
\label{sec:cost-model}


\begin{table}
  \centering
  \begin{tabular}{lll}
    ResNet    & Throughput & Accuracy \\ \hline
    ResNet-18 & 12,592 & 68.2\% \\
    ResNet-34 & 6,860  & 71.9\% \\
    ResNet-50 & 4,513  & 74.34\%
  \end{tabular}
  \caption{Throughput and top-one accuracy for ResNets of different depths. As
  shown, there is a trade off between accuracy and throughput (i.e.,
  computation).}
  \vspace{-1.5em}
  \label{table:acc-thpt}
\end{table}

\begin{table*}
\centering
\setlength\itemsep{2em}
\small
\begin{tabular}{llll|lll}
  Config. & Preprocessing & DNN execution & Pipelined &
      \sn estimate & \blazeit estimate & \tahoma estimate \\
            & throughput (im/s)        & throughput (im/s)        & throughput (im/s)    &
      (\% error, estimate)  & (\% error, estimate) & (\% error, estimate) \\
  \hline
  Balanced      & 4001 & 4999 & 4056 & \textbf{1.4\%}, 4001 & 23.2\%, 4999   & 44.8\%, 2222 \\
  Preproc-bound & 534  & 4999 & 557  & \textbf{4.1\%}, 534  & 797.5\%, 4999  & 9.3\%, 482 \\
  DNN-bound     & 5876 & 1844 & 1720 & \textbf{7.2\%}, 1844 & \textbf{7.2\%}, 1884 & 22.7\%, 1403
\end{tabular}
\caption{We show measurements of preprocessing, DNN execution, and pipelined
end-to-end DNN inference for three configurations of DNNs and input formats:
balanced, preprocessing-bound, and DNN-execution bound. We measure the
throughput in images per second of preprocessing, DNN execution, and end-to-end
DNN inference on the left. We show the throughput estimation and error
in estimation for three cost models on the right. We bold the most accurate estimate. As shown,
\sn matches or ties the most accurate estimate for all conditions.\protect\footnotemark}
\label{table:cost-models}
\vspace{-1.5em}
\end{table*}

When deploying DNN-based visual analytics systems, application developers have
different resource constraints. As such, these systems often expose a way of
trading off between accuracy and throughput. Higher accuracy DNNs typically
require more computation: we demonstrate this property on the popular ImageNet
dataset~\cite{deng2009imagenet} with standard ResNets in
Table~\ref{table:acc-thpt}. Prior work has designed high throughput specialized
DNNs for filtering~\cite{kang2017noscope, anderson2018predicate,
kang2019blazeit}. We do not focus on the design of DNNs in this work and instead
use standard DNNs (\S\ref{sec:acc-thpt}).

One popular method for DNN selection is to use a cost
model~\cite{kang2017noscope, anderson2018predicate}. We describe cost modeling
for DNNs and how prior work estimated the throughput of DNN
execution. Critically, these prior cost models ignore preprocessing costs or
ignore that preprocessing can be pipelined with DNN execution. We show that
ignoring these factors can lead to inaccurate throughput estimations
(Table~\ref{table:cost-models}). We then describe how to make cost models
preprocessing-aware.

\minihead{Cost models}
Given a set of resource constraints and metrics to optimize, a system must
choose which DNNs to deploy to maximize these metrics while respecting resource
constraints. For example, one popular constraint is a minimum throughput and one
popular metric is accuracy. In this exposition, we focus on
throughput-constrained accuracy and accuracy-constrained throughput, but other
constraints could be used.

Specifically, denote the possible set of system configurations as $C_1, ...,
C_n$. Denote the resource consumption estimate of each configuration as $R(C)$
and the resource constraint as $R_{\max}$. Denote the metric to optimize as
$M(C)$.

In its full generality, the optimization problem is
\begin{equation}
\begin{aligned}
\max_i {}  &  M(C_i) \\
\textnormal{s.t. }  &  R(C_i) \leq R_{\max}.
\end{aligned}
\end{equation}
In this framework, both accuracy and throughput can either be constraints or
metrics. For example, for throughput-constrained accuracy, $R(C_i)$ would be an
estimate of the throughput of $C_i$ and $M(C_i)$ would be an estimate of the
accuracy of $C_i$. Similarly, for accuracy-constrained throughput, $R(C_i)$
would be an estimate of the accuracy and $M(C_i)$ would be an estimate of the
throughput.

As an example, \tahoma generates $C_i = [D_{i, 1}, ..., D_{i, k}]$ to be a
sequence of $k$ models, $D_{i, j}$, that are executed in sequence. The resource
$R(C_i) = A(C_i)$ is the accuracy of configuration $C_i$ and the metric $M(C_i)
= T(C_i)$ is the throughput of configuration $C_i$.

Prior work has focused on expanding the set of $C_i$ or evaluating $R(C_i)$ and
$M(C_i)$ efficiently~\cite{kang2017noscope, lu2018accelerating,
anderson2018predicate, canel2019scaling}. A common technique is to use a smaller
model (e.g., a specialized NN) to filter data before executing a larger, target
DNN in a \emph{cascade}. For example, when detecting cars in a video, \noscope
will train an efficient model to filter out frames without
cars~\cite{kang2017noscope}. Cascades can significantly expand the feasible set
of configurations.

For cost models to be effective, the accuracy and throughput measurements must
be accurate. We discuss throughput estimation below. Accuracy can be estimated
using best practices from statistics and machine learning. A popular method is
to use a held-out validation set to estimate the
accuracy~\cite{bishop2006pattern}. Under the assumption that the test set is
from the same distribution as the validation set, this procedure will give an
estimate of the accuracy on the test set.

\minihead{Throughput estimation}
A critical component of cost model for DNNs is the throughput estimation of a
given system configuration $C_i$; recall that $C_i$ is
represented as a sequence of one or more DNNs, $D_{i, j}$. Given a specific DNN
$D_{i,j}$, estimating its throughput simply corresponds to executing the
computation graph on the accelerator and measuring its throughput. As DNN
computation graphs are typically fixed, this process is efficient and accurate.

\miniheadit{Estimation ignoring preprocessing}
Prior work (e.g., probablistic predicates, \blazeit,
\noscope)~\cite{kang2017noscope, kang2019blazeit, lu2018accelerating} has used
the throughput of $D_{i,j}$ to estimate the throughput of end-to-end DNN
inference. Specifically, \blazeit and \noscope estimates the throughput,
$\hat{T}(C_i)$ as
\begin{equation}
  \hat{T}(C_i) \approx \frac
      {1}
      {
        \sum_{j=1}^{k} \frac{1}{\alpha_j^{-1} T_{\mathrm{exec}}(D_{i,j})}
      }
\end{equation}
where $\alpha_j$ is the pass-through rate of DNN $D_{i, j}$ and
$T_{\mathrm{exec}}(D_{i,j})$ is the throughput of executing $D_{i,j}$.
$T_{\mathrm{exec}}(D_{i,j})$ can be directly measured using synthetic data and
$\alpha_j$ can be estimated with a validation set. This
approximation holds when the cost of preprocessing is small compared to the cost
of executing the DNNs.

However, this cost model ignores preprocessing costs. As a result, it is
inaccurate when preprocessing costs dominate DNN execution costs or when
preprocessing costs are approximately balanced with DNN execution costs
(Table~\ref{table:cost-models}).

  \footnotetext{Preprocessing having lower throughput than both in the
  preprocessing-bound and balanced conditions are due to the experimental harness
  being optimized for pipelined execution. The experimental harness does not
  significantly affect throughput when compared to without the harness.}

\miniheadit{Estimation ignoring pipelining}
Other systems (e.g., \tahoma) \cite{anderson2018predicate} estimate end-to-end
DNN inference throughput as
\begin{equation}
  \hat{T}(C_i) \approx \frac{1}{
    \frac{1}{T_{\mathrm{preproc}}(C_i)} +
        \frac{1}{T_{\mathrm{exec}}(C_i)}}.
\end{equation}
This approximation ignores that preprocessing can be pipelined with DNN
execution. As a result, this approximation holds when either preprocessing or
DNN execution is the overwhelming bottleneck, but is inaccurate for other
conditions, namely when preprocessing costs are approximately balanced with DNN
execution costs (Table~\ref{table:cost-models}).

\vspace{0.5em}

These throughput approximations (that ignore preprocessing costs and ignore
pipelining) ignore two critical factors: 1) that input preprocessing can
dominate inference times and 2) that input preprocessing can be pipelined
with DNN execution on accelerators. We now describe a more accurate throughput
estimation scheme.

\minihead{Corrected throughput estimation}
For high throughput DNN inference on accelerators, the DNN execution and
preprocessing of data can be pipelined. As a result, \sn uses a more accurate
throughput estimate for a given configuration:
\begin{equation}
  \hat{T}(C_i) \approx \min \left(
      T_{\mathrm{preproc}}(C_i),
      \frac{1}{
          \sum_{j=1}^{k} \frac{1}{\alpha_j^{-1} T_{\mathrm{exec}}(D_{i, j}) }
      }
  \right)
\end{equation}
Importantly, as we have shown in \S\ref{sec:benchmarks}, preprocessing can
dominate end-to-end DNN inference. While there are some overheads in pipelining
computation, we empirically verify the $\min$ approximation
(\S\ref{sec:eval-im}).

If preprocessing costs are fixed, then it becomes optimal to maximize the
accuracy of the DNN subject to the preprocessing throughput. Namely, the goal is
to pipeline the computation as effectively as possible. We give two examples of
how this can change which configuration is chosen.


First, when correctly accounting for preprocessing costs in a
throughput-constrained accuracy deployment, it is not useful to select a
throughput constraint higher than the throughput of preprocessing. Second, for
an accuracy-constrained throughput deployment, the most accurate DNN subject to
the preprocessing throughput should be selected.

\section{Input-aware Methods for Accuracy and Throughput Trade Offs}
\label{sec:acc-thpt}

Given the corrected cost model, \sn's goal is to maximize the minimum of the
preprocessing and DNN execution throughputs. However, if the input format and
resolution are fixed, preprocessing throughputs are fixed and can be lower than
DNN execution throughputs.

To provide better accuracy and throughput trade-offs, we propose three
techniques: 1) expanding the search space of specialized DNNs, 2)
using natively present, low resolution visual data, and 3) a DNN training
technique to recover accuracy loss from naively using low resolution visual
data.

\subsection{Expanding search space}

As described, many systems only consider cheap, specialized NNs. Concretely,
\blazeit and \tahoma considers specialized NNs that can execute up to 250,000
images/second, which far exceeds preprocessing throughputs for standard image
and video encodings. As DNNs are generally more accurate as they become more
expensive, these systems use specialized NNs that are less accurate relative to
preprocessing throughput-matched NNs.

In contrast, \sn considers NNs that have been historically considered expensive.
We have found that standard ResNet configurations~\cite{he2016deep} (18 to 152)
strongly outperforms specialized NNs used in prior work.  Furthermore, ResNet-18
can execute at 12.6k images/second, which generally exceeds the throughput of
preprocessing.
Thus, \sn currently uses these ResNets as the specialized NNs. As hardware
advances, other architectures (e.g., ResNeXt~\cite{xie2017aggregated}) may be
appropriate.

\subsection{Low-resolution data}
\label{sec:low-res}

\minihead{Overview}
Many visual data services store the data at a range of resolutions.
Low-resolution visual data is typically stored for previewing purposes or for
low-bandwidth situations. For example, Instagram stores 161x161 previews of
images~\cite{arens2019always}. Similarly, YouTube stores several resolutions of
the same video for different bandwidth requirements, e.g., 240p up to 4K video.

Decoding low-resolution visual data is more efficient than decoding full
resolution data. \sn could decode and then upscale the low-resolution visual
data for improved preprocessing throughput.
However, we show that naively upscaling gives low accuracy results. Instead,
\sn will train DNNs to be aware of low-resolution data, as described below
(\S\ref{sec:low-res-training}).

Recent work uses lower resolution data to improve NN throughput,
\emph{but not to reduce preprocessing costs}~\cite{zhang2017live,
anderson2018predicate}. These systems decode full-resolution data and
downsamples the data, which does not improve preprocessing throughput.

\minihead{Selecting DNNs and resolution jointly}
Many systems provide accuracy and throughput trade-offs by cascading a
specialized NN and a more accurate, target DNN~\cite{kang2017noscope,
anderson2018predicate, lu2018accelerating}. However, these
specialized NNs are often bottlenecked by preprocessing costs.

Instead, \sn uses low resolution data reduce preprocessing costs,
and therefore end-to-end execution costs. However, low resolution visual data
discards visual information and can result in lower accuracy in many cases.
Nonetheless, \sn can provide accuracy and throughput trade-offs by carefully
selecting DNN and input format combinations.

As a motivating example, consider ResNet-34 and 50 as the DNNs, and full
resolution and 161x161 PNG thumbnails as the input formats. ResNet-34 and
ResNet-50 execute at 6,861 and 4,513 images/second. On full resolution data,
they achieve 72.72\% and 75.16\% accuracy on ImageNet, respectively. On low
resolution data, they achieve 72.50\% and 75.00\% accuracy, respectively (when
upscaling the inputs to 224x224 and using \sn's augmented training procedure).
Full resolution and 161x161 thumbnails decode at 527 and 1,995 images/second,
respectively. In this example, executing ResNet-50 on 161x161 thumbnails
outperforms executing ResNet-34 on full resolution data, as end-to-end execution
is bottlenecked by preprocessing costs.

Thus, \sn jointly considers both the input resolution format and the DNN. For
classification, \sn also considers using a \emph{single} DNN for
accuracy/throughput trade-offs, instead of cascading a specialized DNN and
target DNN.

For a given input format, \sn will only consider DNNs that exceed the throughput
of the preprocessing costs and select the highest accuracy DNN subject to this
constraint. As we have demonstrated, in certain cases, this will result in
selecting lower resolution data with more expensive DNNs, contrary to prior
work.

\subsection{Training DNNs for Low-resolution Visual Data}
\label{sec:low-res-training}

As described above, \sn can use low-resolution visual data to decrease
preprocessing costs. However, naively using low-resolution can decrease
accuracy, especially for target DNNs. For example, using a standard ResNet-50
with native 161x161 images upscaled to the standard 224x224 input resolution
results in a \textit{10.8\% absolute drop in accuracy}. This drop in accuracy is
larger than switching from a ResNet-50 to a ResNet-18, i.e., nearly reducing the
depth by a third. To alleviate the drop in accuracy, \sn can train DNNs to be
aware of low-resolution. This procedure can recover, or even exceed, the
accuracy of standard DNNs.

\sn trains DNNs to be aware of low-resolution by augmenting the input data at
training time. At training time, \sn will downsample the full-resolution inputs
to the desired resolution and then upsample them to the DNN input resolution.
\sn will do this augmentation in addition to standard data augmentation. By
purposefully introducing downsampling artifacts, these DNNs can be trained to
recover high accuracy on low-resolution data.

We show that this training procedure can recover the accuracy of full resolution
DNNs when using lossless low-resolution data, e.g., PNG compression. However,
when using lossy low-resolution data, e.g., JPEG compression, low-resolution
DNNs can suffer a drop in accuracy. Nonetheless, we show that using lossy
low-resolution data can be more efficient than using smaller, full-resolution
DNNs.

\section{An Optimized Runtime Engine for End-to-End Visual Inference}
\label{sec:preproc-opt}

In order to efficiently execute \emph{end-to-end} visual inference in the
high-throughput setting, we must make proper use of all available hardware. We
describe how to efficiently pipeline preprocessing and DNN execution for full
use of hardware resources, how to optimize common preprocessing operations, how
to place operations on CPUs or accelerators, and methods of partially decoding
visual data. Several of these optimizations have been explored in other
contexts, but not for end-to-end DNN inference~\cite{gale2017high,
eliuk2016dmath}.

\subsection{Efficient Use of Hardware}
In order to efficiently use all available hardware resources, \sn must
efficiently pipeline computation, use threads, and use/reuse memory.

As executing DNNs requires computation on the CPU and accelerator, \sn must
overlap the computation. To do this, \sn uses a multi-producer, multi-consumer
(MPMC) queuing system to allow for multithreading. The producers decode the visual
data and the consumers perform DNN execution. \sn uses multiple consumers to
leverage multiple CUDA streams. As preprocessing is data parallel
and issuing CUDA kernels is low overhead, we find that setting the number of
producers to be equal to the number of vCPU cores to be an efficient heuristic
for non-NUMA servers.

An important performance optimization to effectively use the MPMC queuing system
is reusing memory and efficient copying to the accelerator. Prior work that
focuses on efficient preprocessing for training must pass memory buffers
that contain the preprocessed images to the caller, which does not
allow for efficient memory reuse. In contrast, the caller to \sn only requires
the result of inference, not the intermediate preprocessed buffer. As a
result, \sn can reuse these buffers. Furthermore, accelerators require pinned
memory for efficient memory transfer. Reusing pinned memory results in
substantially improved performance.  \sn will further over-allocate memory to
ensure that producer threads will not contend on consumers.

\subsection{Optimizing Common Preprocessing Operations}
\label{sec:opt-common}

A large class of common visual DNN preprocessing operations fall under the steps
described in \S\ref{sec:benchmarks}. Briefly, they include resizing,
cropping, pixel-level normalization, data type conversion, and channel
reordering. We can optimize these operations at inference time by fusing,
reordering, and pre-computing operations.

To optimize these steps, \sn will accept the preprocessing steps as a
computation directed, acyclic graph (DAG) and performs a combination of
rule-based and cost-based optimization of these steps. To optimize a
computation DAG, \sn will exhaustively generate possible execution plans,
apply rule-based optimization to filter out plans, and perform cost-based
optimization to select between the remaining plans.

\sn contains rules of allowed operation reordering to generate the
possible set of execution plans:
\begin{enumerate}[itemsep=0em,parsep=0em,topsep=0em]
  \item Normalization and data type conversion can be placed at any point in the
  computation graph.
  \item Normalization, data type conversion, and channel reordering can be
  fused.
  \item Resizing and cropping can be swapped.
\end{enumerate}

Once \sn generates all possible execution plans, \sn will then apply the
following rules to prune plans:
\begin{enumerate}[itemsep=0em,parsep=0em,topsep=0em]
  \item Resizing is cheaper with fewer pixels.
  \item Resizing is cheaper with smaller data types (e.g., \texttt{INT8}
  resizing is cheaper than \texttt{FLOAT32} resizing).
  \item Fusion always improves performance.
\end{enumerate}
We currently implement fusion manually, but code generation could also be
applied to generate these kernels~\cite{palkar2017weld}.
Given a set of plans after rule-based pruning, \sn approximates the
cost by counting the number of arithmetic operations in each plan for the given
data types. \sn will select the cheapest plan.


\subsection{Preprocessing Operator Placement}
In addition to optimizing common preprocessing operations, \sn can
place preprocessing operations on the CPU or accelerator. Depending on the input
format/resolution and DNN, the relative costs of preprocessing and DNN execution
may differ. For example, small specialized NNs may execute many times faster
than preprocessing, but a state-of-the-art Mask R-CNN may execute slower than
preprocessing.

As a result, to balance preprocessing and DNN execution costs, it may be
beneficial to place operations on either the CPU or accelerator.  Furthermore,
many preprocessing operations (e.g., resizing, normalization) are efficient on
accelerators, as the computational patterns are similar to common DNN
operations.

If DNN execution dominates, then \sn will place as many operations on the CPU as
possible, to balance costs. If preprocessing cost dominate, then \sn will place
as many operations on the accelerator as possible. Since preprocessing
operations are sequential, \sn need only consider a small number (typically
under 5) configurations for a given model and image format.

\subsection{Partial and Low-Fidelity Decoding}
\label{sec:partial-decoding}


\minihead{Overview of Visual Compression Formats}
We briefly describe salient properties of the majority of popular visual
compression formats, including the popular JPEG, HEVC/HEIC, and H.264
compression formats. We describe the decoding of the data and defer a
description of encoding to other texts~\cite{pennebaker1992jpeg,
sullivan2012overview, wiegand2003overview}.

Decoding generally follows three steps: 1) entropy decoding, 2) inverse
transform (typically DCT-based), and 3) optional post-processing for improved
visual fidelity (e.g., deblocking).

Importantly, the entropy decoders in both JPEG and HEVC (Huffman decoding and
arithmetic decoding respectively) are not efficient on accelerators for DNNs as
it requires substantial branching. Furthermore, certain parts of decoding can be
omitted, e.g., the deblocking filter, for reduced fidelity but faster decoding
times.

%
%

\begin{figure}
  \includegraphics[width=\columnwidth]{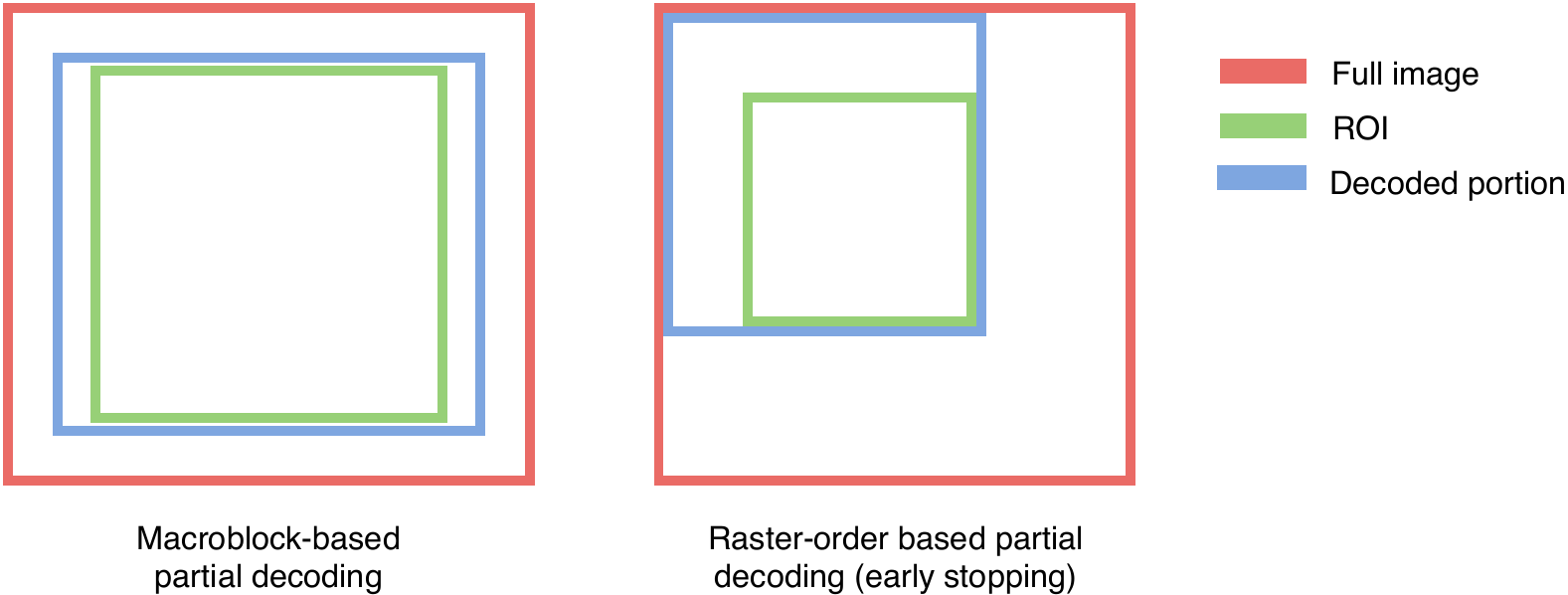}
  \vspace{-0.5em}
  \caption{Examples of partial decoding for images. On the left, the ROI is the
  central crop of the image. For JPEG images, \sn can decode only the
  macroblocks that intersect the ROI. For image formats that do not allow for
  independently decoding macroblocks, \sn can partially decode based on raster
  order (right). Thus, only the blue portion need be decoded. As decoding is
  generally more expensive than other parts of the preprocessing pipeline,
  partial decoding can significantly improve throughput.}
  \label{fig:decoding}
  \vspace{-1.5em}
\end{figure}

\minihead{Leveraging partial decoding}
When low-resolution visual data is not available, \sn can optimize preprocessing
by partially decoding visual data. Many DNNs only require a portion of the image
for inference, or \emph{regions of interest (ROI)}. For example, many image
classification networks centrally crop images, so the ROI is the central crop.
Computing face embeddings crops faces from the images, the ROIs are the face
crops. Furthermore, these networks often take standard image sizes, e.g.,
$224\times224$ for the standard ResNet-50. We show two examples in
Figure~\ref{fig:decoding}. \colora{Computing ROIs may require expensive
upstreaming processing in some applications, e.g., executing a detection DNN.}

Many image compression formats allow for partial decoding explicitly in the
compression standard and all compression formats we are aware of allow for early
stopping of decoding. We give three instantiations of partial decoding in
popular visual compression formats and provide a list of popular visual data
compression formats and which features they contain in
Table~\ref{table:compression-features}. We then describe how to use these
decoding features for optimized preprocessing.

First, for the JPEG image compression standard, each 8x8 block, or
\emph{macroblock}, in the image can
be decoded independently (partial decoding) \cite{wallace1992jpeg}. Second, the
H.264 and HEVC video codecs contain deblocking filters, which can
be turned off at the decoding stage for reduced computational complexity at the
cost of visual fidelity (reduced fidelity decoding)~\cite{sullivan2012overview,
wiegand2003overview}. Third, the JPEG2000 image compression format contains
``progressive'' images, i.e., downsampled versions of the same image, that can
be partially decoded to a specific resolution (multi-resolution
decoding)~\cite{taubman2012jpeg2000}.

\sn accepts as an optional input an ROI for a given image. If an ROI is
specified, \sn will only decode the parts of the image necessary to process the
ROI.

\minihead{Partial decoding}
We present two methods of partially decoding visual data. We show examples of
each in Figure~\ref{fig:decoding}.

\miniheadit{ROI decoding}
When only a portion of the image is needed, e.g., for central cropping or when
selecting a region of interest (ROI), only the specified portion of the image
need be decoded. To decode this portion of the image, \sn will first find the
smallest rectangle that aligns with the 8x8 macroblock border and contains the
region. Then, \sn will decode the rectangle and return the crop. This procedure
is formalized in Algorithm~\ref{alg:partial}.

\miniheadit{Early stopping}
For compression formats that do not explicitly allow for partial decoding, \sn
can terminate decoding on parts of the image that are not necessary. For
example, if only the top $N \times N$ pixels are required for inference, \sn
will terminate decoding after decoding the top $N \times N$ pixels.

\begin{algorithm}[t!]
\begin{algorithmic}
  \ENSURE $img \in (3, h, w)$
  \STATE $ w', h' \leftarrow RatioPreservingResize(img) $
  \STATE $l, t = \frac{w' - 224}{2}, \frac{h' - 224}{2}$
  \STATE $r, b = l + 224, t + 224$
  \STATE $scale \leftarrow \frac{\min(h, w)}{224}$
  \STATE $l', r', t', b' = scale \cdot [l, r, t, b]$
  \STATE img.SetCropline($l'$, $r'$, $t'$)
  \WHILE{rowIdx $\leq b'$}
    \STATE $img[$rowIdx$]$ = ReadScanlines(rowIdx)
    \STATE rowIdx $+= 1$
  \ENDWHILE
\end{algorithmic}
\caption{Partial JPEG decoding for a fixed DNN input resolution of
$224\times224$}
\label{alg:partial}
\end{algorithm}

\minihead{Reduced-fidelity decoding}
Several visual compression formats contain options for reduced fidelity
decoding. While there are several ways to reduce the fidelity of decoding for
decreased preprocessing costs, we focus on methods that are easily specified
with existing decoding APIs. Specifically, we explore reduced fidelity in the
form of disabling the deblocking filter. \sn will profile the accuracy of the
specialized and target NNs with and without the deblocking filter and choose the
option that maximizes throughput.


\begin{table}
\centering
\setlength\itemsep{2em}
\begin{tabular}{lll}
Format    & Type  & Low-fidelity features \\ \hline
JPEG      & Image & Partial decoding \\
PNG, WebP & Image & Early stopping \\
HEIC/HEVC & Image/Video & Reduced fidelity decoding \\
H.264     & Video & Reduced fidelity decoding \\
VP8       & Video & Reduced fidelity decoding \\
VP9       & Video & Reduced fidelity decoding
\end{tabular}
\caption{A list of popular visual data formats and their low-fidelity features.
Many popular formats contain methods of decoding parts of the visual data,
including the popular JPEG, H.264, and HEVC formats.}
\label{table:compression-features}
\vspace{-1.5em}
\end{table}


\section{Discussion on Hardware and Power}
\label{apx:hardware}

\minihead{Overview}
Throughout, we use the \texttt{g4dn.xlarge} instance as our testing environment.
The \texttt{g4dn.xlarge} instance has a single NVIDIA T4 GPU, 4 vCPU cores
(which are hyperthreads), and 15GB of RAM. The CPU type is the Intel Xeon
Platinum 8259CL CPU, which is a proprietary CPU developed specifically for this
instance type. Its power draw is 210 watts, or 4.375 watts per vCPU
core.

\minihead{Discussion}
We note that there are other \texttt{g4dn} instances which contain a single T4
GPU and 8, 16, 32, and 64 vCPU cores. These other instances types could be used
to improve the preprocessing throughput by using more cores.

Nonetheless, our speedup numbers can be converted to cost savings when
considering other instance types and our conclusions remain unchanged with
respect to cost. For example, a $3\times$ improvement in throughput of
preprocessing can be translated to using $3\times$ fewer cores.  Furthermore,
the cost of additional cores dominates: around 3.4 vCPU cores is the same price
as the T4 when estimating the cost of vCPU cores and the T4 (see below).

Using these price and power estimates, we estimate the relative price and
power of preprocessing and DNN execution. Preprocessing is significantly more
expensive than DNN execution for the ResNet-50, both in terms of cost (\$0.218
vs \$2.37 per hour) and power draw (70W vs 161W), for the configuration in
Figure~\ref{fig:bs64-naive}. For ResNet-18, these
differences are more prominent: \$0.218 vs \$6.501 and 444W vs 70W for price and
power respectively.

\begin{table}
\centering
\setlength\itemsep{2em}
\begin{tabular}{lll}
GPU  & Release date & Throughput (im/s) \\ \hline
K80  & 2014         & 159   \\
P100 & 2016         & 1,955 \\
T4   & 2019         & 4,513 \\
V100 & 2017         & 7,151 \\
RTX  & 2019         & 15,008 (reported)
\end{tabular}
\caption{Throughput of ResNet-50 on GPU accelerators. Throughput has
improved by over 94$\times$ in three years and will continue to improve. The T4
is an inference optimized accelerator that is significantly more power efficient
than the V100, but contains similar hardware units.}
\label{table:gpu-throughput}
\vspace{-1.5em}
\end{table}

\minihead{Core price estimation}
We estimate the price per vCPU core using a linear interpolation, assuming the
T4 is a fixed price and the remaining price is split equally among the cores.
Using this method, we find that the hourly cost of the T4 accelerator is
approximately \$0.218 and the cost of a single vCPU core is approximately
\$0.0639. The $R^2$ value of this fit is 0.999. Thus, approximately 3.4 vCPU
cores is the same hourly price of a T4.

\minihead{Trends in Hardware Acceleration for DNNs}
We benchmarked ResNet-50 throughput on the K80, P100, T4, and V100 GPUs to show
the effect of improved accelerators on throughput; we further show the reported
throughput of an unreleased accelerator~\cite{wong2018habana}. We used a batch
size of 64 for experiments on GPUs. As shown in
Table~\ref{table:gpu-throughput}, throughput has improved by 44$\times$ in three
years. Furthermore, accelerators will become more efficient.


%

\section{Evaluation}
\label{sec:eval}

We evaluated \sn on eight visual datasets and show that \sn can outperform
baselines by up to 5.9$\times$ for image datasets and 10$\times$ for video
datasets at a fixed accuracy level.

\subsection{Experimental Setup}

\minihead{Overview}
We evaluate our optimizations on four image datasets and four video datasets.
The task for the image datasets is image classification. The task for the video
datasets is an aggregation query for the number of target objects per frame. For
classification, we use accuracy and throughput as our primary evaluation
metrics. For the aggregation queries, we measure query runtime as the error
bounds were respected.

\colora{We describe \sn's implementation in an extended version of this paper.}

\begin{table}
\centering
\setlength\itemsep{2em}
\begin{tabular}{llll}
  Dataset & \# of classes & \# of train im. & \# of test im. \\ \hline
  \texttt{bike-bird}  & 2     & 23k   & 1k \\
  \texttt{animals-10} & 10    & 25.4k & 2.8k \\
  \texttt{birds-200}  & 200   & 6k    & 5.8k \\
  \texttt{imagenet}   & 1,000 & 1.2M  & 50K
\end{tabular}
\caption{Summary of dataset statistics for the still image datasets we used in
our evaluation. The datasets range in difficulty and number of classes.
\texttt{bike-bird} is the easiest dataset to classify and \texttt{imagenet} is
the hardest to classify.}
\label{table:image-dataset-stats}
\vspace{-1.5em}
\end{table}

\minihead{Datasets}
We use \texttt{bike-bird} \cite{brown2018unrestricted}, \texttt{animals-10}
\cite{alessio2019animals}, \texttt{birds-200} \cite{wah2011caltech}, and
\texttt{imagenet} \cite{deng2009imagenet} as our image datasets. These datasets
vary in difficulty and number of classes (2 to 1,000). In contrast, several
recent systems study only binary filtering \cite{anderson2018predicate,
canel2019scaling, lu2018accelerating}. We summarize dataset statistics in
Table~\ref{table:image-dataset-stats}. We used thumbnails encoded in a standard
short size of 161 in PNG, JPEG ($q=75$), and JPEG ($q=95$).

For the video datasets, we used \texttt{night-street}, \texttt{taipei},
\texttt{amsterdam}, and \texttt{rialto} as evaluated by
\blazeit~\cite{kang2019blazeit}. We used the original videos as evaluated by
\blazeit and further encoded the videos to 480p for the low-resolution versions.

\minihead{Model configuration and baselines}
For \sn, we use the standard configurations of ResNets, specifically 18, 34, and
50. We find that these models span a range of accuracy and speed while only
requiring training three models. We note that if further computational resources
are available at training time, further models could be explored.

\miniheadit{Image datasets}
For the image datasets, we use the following two baselines. First, we use
standard ResNets and vary their depths, specifically choosing 18, 34, and 50 as
these are the standard configurations~\cite{he2016deep}. We refer to this
configuration as the naive baseline; the naive baseline does not have access to
other image formats. Second, we use \tahoma as our other baseline, specifically
a representative set of 8 models from \tahoma cascaded with ResNet-50, our most
accurate model. We choose 8 models due to the computational cost of training
these models, which can take up to thousands of GPU hours for the full set of
models. \colora{We use ROI decoding for \sn as these datasets use central crops.}

\miniheadit{Video datasets}
We used the original \blazeit code, which uses a ``tiny ResNet'' as the
specialized NN and a state-of-the-art Mask R-CNN~\cite{he2017mask} and
FGFA~\cite{zhu2017flow} as target networks. We replicate the exact experimental
conditions of \blazeit, except we use \sn's optimized runtime engine, which is
substantially more efficient than \blazeit.

\minihead{Hardware environment}
We use the AWS \texttt{g4dn.xlarge} instance type with a single NVIDIA T4 GPU
attached unless otherwise noted. The \texttt{g4dn.xlarge} has 4 vCPU cores with
15 GB of RAM. A vCPU is a hyperthread, so 4 vCPUs consists of 2
physical cores. Compute intensive workloads, such as image decoding,
will achieve sublinear scaling compared to a single hyperthread.

Importantly, the \texttt{g4dn.xlarge} instance is approximately cost balanced
between vCPU cores and the accelerator (\S\ref{apx:hardware}).

\texttt{g4dn.xlarge} is optimized for DNN inference. Namely, the
T4 GPU is significantly more power efficient than GPUs designed for training,
e.g., the V100. However, they achieve lower throughput as a result; our results
are more pronounced when using the V100 (e.g., using the \texttt{p3.2xlarge}
instance). We further describe our choice of hardware environment in
\S\ref{apx:hardware}. We note that our baselines use CPU decoding as a
case study, as not all visual formats are supported by hardware decoders, e.g.,
the popular HEIC (used by all new iPhones) and WebP (used by Google Chrome)
formats. Finally, we note that throughputs can be converted to power or cost by
using more vCPU cores, but we use a single hardware environment for ease of
comparison.

\minihead{Further experiments}
\colora{Due to limited space, we include experiments comparing \sn to other
frameworks in an extended version of this paper.}

\subsection{Cost Models and Benchmarking \sn}

We further investigated the efficiency of pipelining in \sn and our choice of
using $\min$ in cost modeling (\S\ref{sec:cost-model}). To study these, we
measured the throughput of \sn when only preprocessing, only executing the DNN
computational graph, and when pipelining both stages.

We first consider low-resolution images encoded in JPEG $q=75$ to ensure the
system was under full load. Preprocessing, DNN execution, and end-to-end
inference achieve 5.9k, 4.2k, and 3.6k im/s respectively. Even at full load,
\sn only incurs a 16\% overhead compared to the throughput predicted by its cost
model. In contrast, \tahoma's cost model would predict a throughput of 2.5k
im/s, a 30\% error.

Furthermore, across all ResNet-50 configurations, \sn's cost model (i.e., $\min$)
achieves the lowest error compared to other heuristics (i.e., DNN execution only
and sum). Its average error is 5.9\%, compared to 217\% (DNN execution only) and
23\% (sum).


\subsection{Image Analytics Experiments}
\label{sec:eval-im}

\begin{figure}[t!]
  \includegraphics[width=\columnwidth]{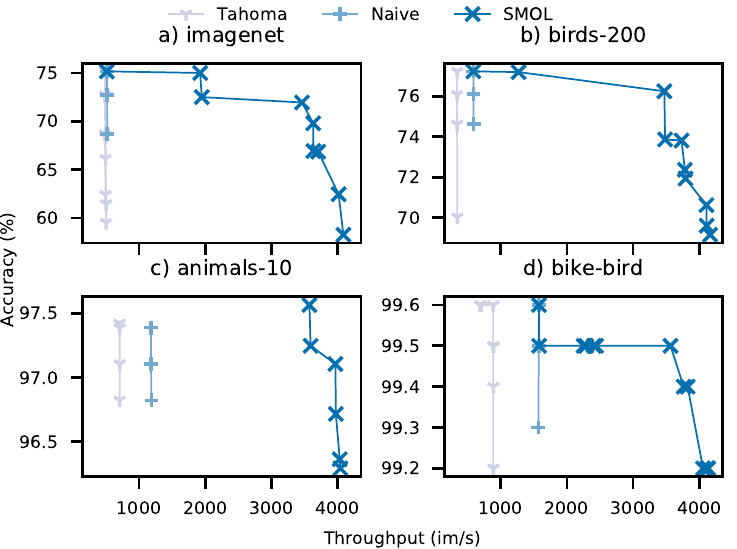}
  \vspace{-0.5em}
  \caption{Throughput vs accuracy for the naive baseline, \tahoma, and \sn on
  the four image datasets (Pareto frontier only).
  \sn can improve throughput by up to 5.9$\times$ with no loss in accuracy.
  Furthermore, \sn can improve the Pareto frontier compared to both baselines.}
  \label{fig:fu-im}
  \vspace{-1em}
\end{figure}

\minihead{End-to-end speedups}
We evaluated \sn and baselines (\tahoma and standard ResNets on full resolution
data) on the image datasets shown in Table~\ref{table:image-dataset-stats}.

We first investigated whether \sn outperforms baselines when
all optimizations were enabled. We plot in Figure~\ref{fig:fu-im} the Pareto
frontier of baselines and \sn for different input format and DNN configurations.
As shown, \sn can improve throughput by up to 5.9$\times$ with no loss in
accuracy relative to ResNet-18 and up to 2.2$\times$ with no loss in accuracy
relative to ResNet-50. Furthermore, \sn can improve the Pareto frontier compared
to all baselines. Notably, \tahoma's specialized models performs poorly on
complex tasks and are bottlenecked on image preprocessing.

Importantly, we see that the naive baselines (i.e., all ResNet depths) \emph{are
bottlenecked by preprocessing} for all datasets. Any further optimizations to
the DNN execution alone, including model compression, will \emph{not improve
end-to-end throughputs}. The differences in baseline throughputs are due to the
native resolution and encoding of the original datasets: \texttt{birds-200}
contains the largest average size of images. The throughput variation between
ResNets depths is due to noise; the variation is within margin of error.

While we show below that both low resolution data and preprocessing
optimizations contribute to high throughput, we see that \sn's primary source of
speedups depends on the dataset. First, \sn can achieve the same or higher
accuracy by simply using low resolution data for some \texttt{bike-bird} and
\texttt{animals-10}. Second, for \texttt{imagenet}, a fixed model will result in
slightly lower accuracy ($\leq1\%$) when using lossless image compression.
However, when using a \emph{larger} model, \sn can recover accuracy.

\begin{table*}
\centering
\setlength\itemsep{2em}
\small
\begin{tabular}{l|cc|cc}
  Format             & Acc (reg train, 50) & Acc (low-resol train, 50) &
      Acc (reg train, 34) & Acc (low-resol train, 34) \\ \hline
  Full resol         & 75.16\% & 57.72\% & 72.72\% & 64.76\% \\
  161, PNG           & 70.92\% & 75.00\% & 68.30\% & 72.50\% \\
  161, JPEG ($q=95$) & 68.93\% & 71.94\% & 66.92\% & 69.79\% \\
  161, JPEG ($q=75$) & 64.02\% & 63.23\% & 62.45\% & 62.45\%
\end{tabular}
\caption{Effect of training procedure and input format on accuracy for ResNet-50
and ResNet-34 on \texttt{imagenet}, the most difficult dataset. \sn can achieve
an accuracy throughput trade-off by simply changing the input format, e.g.,
low-resolution ResNet-50 (low-resol train, 50) on 161, JPEG ($q=95$) achieves
approximately the same accuracy as ResNet-34 (reg train, 34) on full resolution
data (full resol), namely 71.94\% accuracy compared to 72.72\% accuracy. \sn can
also achieve no loss in accuracy for easier datasets (e.g.,
\texttt{bike-bird}).}
\label{table:acc-training}
\vspace{-1.5em}
\end{table*}

\minihead{Comparison against \tahoma}
\tahoma underperforms the naive solution of using a single, accurate DNN for
preprocessing bound workloads. This is primarily due to overheads in cascades,
namely coalescing and further preprocessing operations. Specifically, \tahoma
cascades a small DNN into a larger DNN. These smaller DNNs are
less accurate than the larger DNNs and thus require many images to be passed
through the cascade for higher accuracy, especially on the more complex tasks.
The images that are passed through must be copied again and further resized if
the input resolutions are different.

\minihead{Factor analyses and lesion studies}
We further investigated the source of speedups of \sn's optimizations by
performing factor analyses and lesion studies.

\begin{figure}[t!]
  \includegraphics[width=\columnwidth]{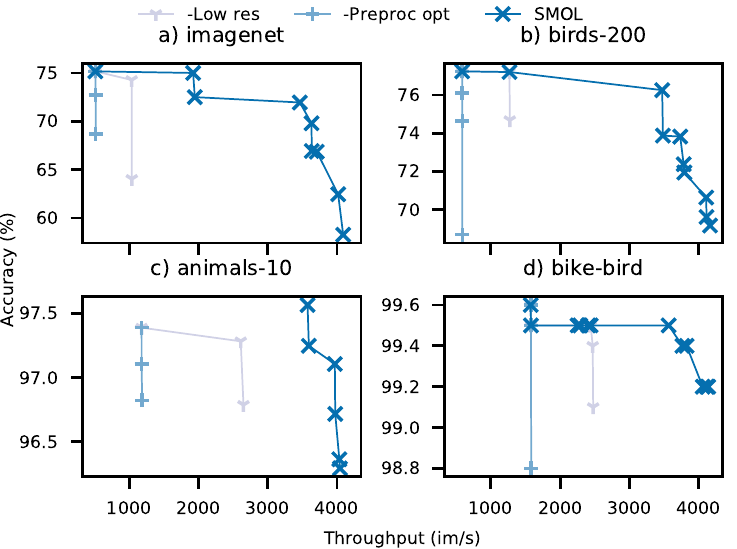}
  \vspace{-0.5em}
  \caption{Lesion study for image datasets in which we individually removed the
  preprocessing optimizations and low-resolution data (Pareto frontier only). As
  shown, both optimizations improve the Pareto frontier for all datasets.}
  \label{fig:lesion-im}
  \vspace{-1em}
\end{figure}

\begin{figure}[t!]
  \includegraphics[width=\columnwidth]{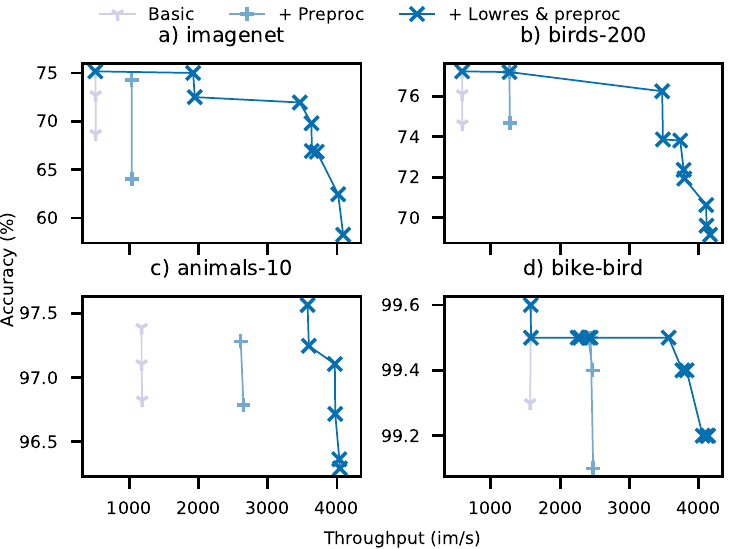}
  \vspace{-0.5em}
  \caption{Factor analysis for image datasets in which we successfully add the
  preprocessing optimizations and then the low-resolution data (Pareto frontier
  only). Both optimizations improve the Pareto frontier for all
  datasets.}
  \label{fig:ablation-im}
  \vspace{-1em}
\end{figure}

We performed a lesion study by individually removing the 1) preprocessing DAG
optimizations and 2) low-resolution data from \sn. As shown in
Figure~\ref{fig:lesion-im}, removing either optimization shifts the Pareto
frontier.

We performed a factor analysis by successively adding the preprocessing DAG and
low-resolution optimizations to \sn. As shown in Figure~\ref{fig:ablation-im},
both optimizations improve throughput. We further see that the optimizations are
task-dependent: the easiest task (\texttt{bike-bird}) can achieve high
throughput at fixed accuracies with only the preprocessing optimizations.
However, many real-world tasks are significantly more complicated than binary
classification of birds and bikes.

\begin{figure}[t!]
  \includegraphics[width=\columnwidth]{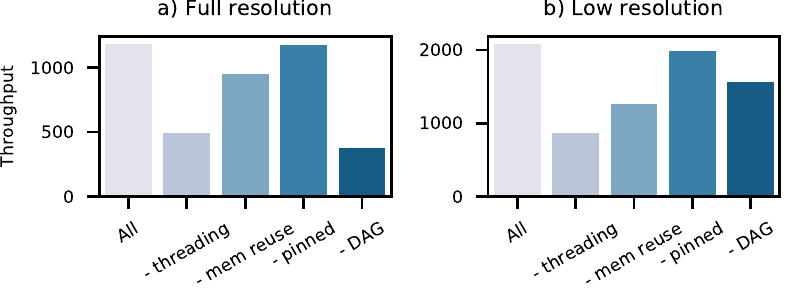}
  \vspace{-0.5em}
  \caption{Lesion study of \sn's systems optimizations for full
  resolution and low resolution images, where optimizations are removed
  individually. All factors contribute to performance. }
  \label{fig:lesion-opt}
  \vspace{-1em}
\end{figure}

\begin{figure}[t!]
  \includegraphics[width=\columnwidth]{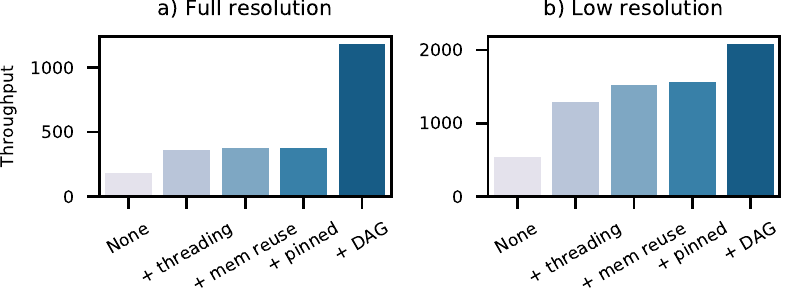}
  \vspace{-1em}
  \caption{Factor analysis of \sn's systems optimizations for full
  resolution and low resolution images, where optimizations are added in
  sequence. All factors contribute to performance.}
  \label{fig:factor-opt}
  \vspace{-1em}
\end{figure}

We also performed a lesion study (Figure~\ref{fig:lesion-opt}) and factor
analysis (Figure~\ref{fig:factor-opt}) for \sn's systems optimizations. We
performed these analyses for ResNet-50 with full resolution and 161 short-side
PNG images on the ImageNet dataset to ensure DNN execution was not the
bottleneck. As shown, all systems optimizations improve performance.
Furthermore, certain optimizations (e.g., DAG vs threading) contribute more to
low resolution and full resolution performance.

\minihead{Effect of training procedure}
We investigated the effect of the training procedure for low-resolution input
formats. We trained ResNet-50 on: 1) full resolution, 2) 161 short-side PNG, 3)
161 short-side JPEG ($q=95$), and 4) 161 short-side JPEG ($q=75$).

We show the accuracy of these conditions in Table~\ref{table:acc-training} for
\texttt{imagenet}, our hardest dataset. As shown, low-resolution aware training
can nearly recover the accuracy of full resolution data even on this difficult
dataset. Low-resolution training can fully recovery accuracy on
\texttt{bike-bird} and \texttt{animals-10}.

\subsection{Video Analytics Experiments}





\begin{figure}[t!]
  \includegraphics[width=\columnwidth]{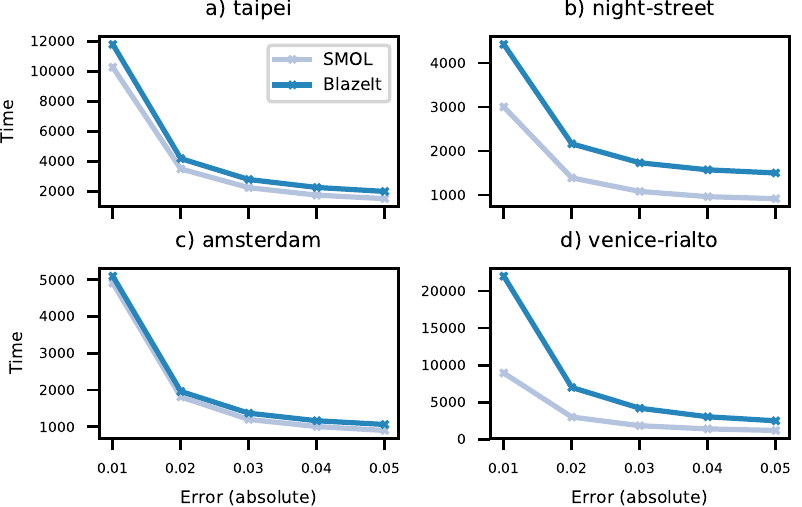}
  \vspace{-0.5em}
  \caption{Query execution time vs requested error for \blazeit and \sn on the
  four video datasets we evaluated. As shown, \sn consistently outperforms
  \blazeit by using more accurate specialized NNs, which reduces sampling
  variance, and lower resolution data, which reduces preprocessing costs.}
  \label{fig:fu-vid}
  \vspace{-0.5em}
\end{figure}

We evaluated \sn on the four video datasets described above. We used the exact
experimental configuration from \blazeit as the baseline, with the exception of
executing \blazeit's specialized NNs in \sn's optimized runtime engine. \sn's
runtime engine is substantially more efficient that \blazeit's.

As shown in Figure~\ref{fig:fu-vid}, \sn can improve throughput by up to
2.5$\times$ at a fixed error level. Furthermore, \sn outperforms \blazeit in all
settings. \sn's primary speedups for \texttt{night-street} and \texttt{rialto}
come from more accurate, but more expensive specialized NNs. Despite the
specialized NNs being more expensive, they reduce sampling variance more, as
they are more accurate. As a result, fewer samples are necessary for a fixed
error target. \sn's primary speedups for \texttt{taipei} and \texttt{amsterdam}
come from leveraging low resolution video, as it is cheaper to preprocess.

\subsection{Cost Analysis}

\begin{table}
\centering
\setlength\itemsep{2em}
\small
\begin{tabular}{llll}
  Condition & vCPUs & Throughput (im/s) & Cost (\textcent / 1M images) \\
  \hline
  Opt    & 4   & 1927 & 7.58 \\
  No opt & 4   & 377  & 38.75 \\
  Opt    & 8   & 3756 & 5.56 \\
  No opt & 8   & 634  & 32.92 \\
  Opt    & 16  & 4548 & 7.35 \\
  No opt & 16  & 1165 & 28.68
\end{tabular}
\caption{
  Throughput and cost of \sn with and without optimizations at variable number
  of vCPU cores to achieve 75\% accuracy on ImageNet. While increasing the
  number of cores improves throughput, \sn's optimizations decrease the
  per-image cost in all cases.
}
\label{table:vcpu-vs-cost}
\end{table}

We analyze how preprocessing affects the cost of inference and how preprocessing
scales with the number of vCPU cores on public cloud instances. We show the
throughput and cost of \sn with and without optimizations to reach an accuracy
of 75\% on ImageNet in Table~\ref{table:vcpu-vs-cost}. We use the prices of the
AWS \texttt{g4dn} instances.
Increasing the number of vCPU cores improves throughput up until matching the
throughput of ResNet-50 on the T4. Nonetheless, \sn is the most cost effective
by up to 5$\times$ per image. We also see that \sn scales nearly linearly
with the number of cores, indicating its efficiency.

\section{Related Work}
\label{sec:related_work}

\minihead{Visual analytics systems}
Contemporary visual analytics systems leverage DNNs for high accuracy
predictions and largely focus on optimizing the cost of executing these
DNNs~\cite{kang2017noscope, anderson2018predicate, canel2019scaling,
kang2019challenges, lu2018accelerating}. These systems typically use smaller
proxy models, such as specialized NNs to accelerate analytics. However, as we
have showed, modern hardware and compilers can create bottlenecks elsewhere in
the end-to-end execution of DNNs.

Other video analytics systems, such as \textsc{Scanner}~\cite{poms2018scanner}
or \textsc{VideoStorm}~\cite{zhang2017live} optimize queries as a black box.
These systems aim to use all available hardware resources but do not jointly
optimize preprocessing and DNN execution.

\minihead{Systems for optimized DNN execution and serving}
Researchers have proposed compilers for optimizing DNN computation graphs,
including TensorRT~\cite{nvidia2019tensort} and others~\cite{leary2017xla,
pytorch2018jit, chen2018tvm}. These compilers generally cannot jointly optimize
preprocessing and DNN execution. Furthermore, as they generate more efficient
code, preprocessing bottlenecks will only increase.

\minihead{Other optimizations for DNN execution}
Researchers have proposed machine learning techniques from model
distillation~\cite{hinton2015distilling} to model compression~\cite{han2015deep}
to reduce the cost of DNN execution. These techniques generally take a given DNN
architecture and improve its accuracy or speed. We are unaware of work
in the machine learning literature for preprocessing-aware optimizations. These
optimizations further improve DNN throughput, but will only increase the gap
between preprocessing and DNN execution.

\minihead{Optimizing DNN preprocessing}
To the best of our knowledge, the only system that focuses on optimizing DNN
preprocessing is NVIDIA DALI~\cite{nvidia2019dali}. However, DALI optimizes
preprocessing for DNN training and focuses on data augmentation. We show that
this results in suboptimal performance in the inference setting.

\minihead{Accelerators for DNNs}
Due to the computational cost of DNN computational graphs, researchers and
companies have created accelerators for DNN execution. These work exclusively
focus on optimizing DNN execution: no paper we surveyed measured
\emph{end-to-end} execution time \cite{jouppi2017datacenter,
fowers2018configurable, han2016eie, chen2016diannao, farabet2011neuflow,
moons20160, chi2016prime, chakradhar2010dynamically, venkataramani2017scaledeep,
li2017drisa, reagen2016minerva, peemen2013memory, park2015energy,
gokhale2014240, chen2016eyeriss, judd2016stripes, albericio2016cnvlutin,
albericio2017bit, parashar2017scnn, shen2017maximizing, rahman2016efficient}.
Until very recently, executing the DNN computational graph was the overwhelming
bottleneck in DNN execution, but we show evidence that trend has reversed
(\S\ref{sec:benchmarks}). Thus, we believe it is critical to reason about
\emph{end-to-end} performance.

\section{Discussion}

We have shown that jointly optimizing preprocessing and DNN
execution can give large improvements in throughput for analytics on existing
visual media compression formats. However, we believe that our techniques can be
further extended as future work.

First, similar techniques can likely be applied to other multimedia. For
example, many audio compression techniques have similar features to visual
compression. Furthermore, audio compression also has a natural trade-off between
fidelity and quality.

Second, as DNN accelerators become more efficient, we believe there is promise
in the joint design between compression and DNN network design. For example, the
massively parallel arithmetic units in DNN accelerators are not well suited to
accelerate standard entropy decoding in existing visual formats. However, they
may be efficient for other forms of compression. Furthermore, compression
algorithms could be designed to improve DNN accuracy.

\section{Conclusion}

In this work, we show that preprocessing can be the bottleneck in end-to-end DNN
inference. We show that the preprocessing costs are accounted for incorrectly in
cost models for selecting models in visual analytics applications. To address
these issues, we build \sn, an optimizing runtime engine for end-to-end DNN inference.
\sn contains two novel optimization for end-to-end DNN inference: 1) an improved
cost model for estimating DNN throughput and 2) joint optimizations for
preprocessing and DNN execution that leverage low-resolution data.  We evaluate \sn
and these optimizations and show that \sn can achieve up
to 5.9$\times$ improvements in throughput.

\begin{acks}
\small
We thank Sahaana Suri, Kexin Rong, and members of the Stanford Infolab for their
feedback on early drafts. This research was supported in part by affiliate
members and other supporters of the Stanford DAWN project---Ant Financial,
Facebook, Google, Infosys, NEC, and VMware---as well as Toyota Research
Institute, Northrop Grumman, Amazon Web Services, Cisco, and the NSF under
CAREER grant CNS-1651570. Any opinions, findings, and conclusions or
recommendations expressed in this material are those of the authors and do not
necessarily reflect the views of the NSF. Toyota Research Institute ("TRI")
provided funds to assist the authors with their research but this article solely
reflects the opinions and conclusions of its authors and not TRI or any other
Toyota entity.
\end{acks}

\balance

\bibliographystyle{ACM-Reference-Format}
\bibliography{vision-preproc}

\clearpage

\begin{appendix}
\section{Implementation}
\label{sec:implementation}

\minihead{Overview}
We implement a \sn prototype with the above optimizations. \sn's training phase is
written in Python and PyTorch. We ported \tahoma's training phase to
PyTorch and used \blazeit's original training code where possible (\sn expands
the set of specialized NNs, so we modified \blazeit's code to accommodate this
change). The execution engine is written in C++.

\minihead{Components}
\sn consists of three major components: 1) data ingestion, 2) preprocessing, and
3) DNN execution components. We describe each component in turn.

\sn provides an API to accept visual data (i.e., images or video) and
a default method for ingesting visual data from disk. Currently, \sn supports
JPEG compressed images or H.264 compressed video due to their popularity. \sn
could be easily extended to a wide range of formats and similar techniques would
apply. For example, JPEG2000 has progressively sized
images~\cite{christopoulos2000jpeg2000}, which could easily be integrated with
low-resolution decoding.

\sn provides standard methods for decoding and preprocessing visual data. \sn
currently supports colorspace conversion, resizing, cropping, and color
normalization. These methods cover a large range of visual analytics tasks,
including classification~\cite{he2016deep}, object detection~\cite{liu2016ssd,
he2017mask}, and face detection~\cite{zhang2016joint}. A user of \sn can also
provide user-defined preprocessing methods.

\sn currently accepts DNNs in the form of Open Neural Network Exchange (ONNX)
computation graphs~\cite{onnx2019}. We choose ONNX as it is widely supported,
e.g., PyTorch, TensorFlow, and Keras can both export to ONNX, and TensorRT
accepts ONNX computation graphs. \sn could easily be extended to support other
backends as new standards and hardware emerge.

\minihead{Low-level details}
To achieve high throughput, \sn leverages three core techniques: 1) efficient
memory layout and reuse, 2) lightweight pipelining, and 3) the use of optimized
libraries.

\miniheadit{Memory optimizations}
To optimize memory usage, \sn allocates buffers once and reuses them when
possible. For example, the input to the DNN computational graph generally has
the same shape so \sn will allocate a buffer for a batch of data and reuse it
between successive batches of data. \sn will also pin memory that will be
subsequently transferred to the GPU. Pinning memory substantially improves copy
performance.

\miniheadit{Pipelining}
For lightweight pipelining, \sn uses a thread pool for preprocessing workers and
CUDA streams for parallel execution. To communicate between the preprocessing
workers and the CUDA execution streams, \sn uses a MPMC queue. Additionally, \sn
only passes pointers between workers, avoiding excessive memory copies.

\miniheadit{Optimized libraries}
\sn uses optimized libraries where possible. For its MPMC queue, \sn uses
\texttt{folly}'s \texttt{MPMCQueue} \cite{folly2019}. For its DNN computation
graph compiler, \sn currently uses TensorRT~\cite{nvidia2019tensort}. We have
found TensorRT to be effective on GPUs, but other compilers could be used for
different hardware substrates. Finally, \sn uses \texttt{FFmpeg},
\texttt{libspng}, and \texttt{libturbo-jpeg} for loading video, PNG images, and
JPEG images respectively.

\subsection{Comparison to DALI and PyTorch}

We compare against NVIDIA's DALI library (a fast preprocessing library for DNN
training) with TensorRT and PyTorch. For all experiments, we use
ResNet-50 on ImageNet and official examples for ResNet-50 validation, which is
equivalent to the preprocessing for inference. We further vary the number of
vCPU cores in this section to understand the scaling of DALI and
PyTorch.

DALI was specifically designed for training and integration with Python
libraries, so its officially supported operations can underperform \sn for
inference. Namely, DALI does not officially support memory-efficient integration
with inference libraries, ROI decoding for JPEG images for inference, and
hardware-aware placement of preprocessing operations. We note that these
limitations are not inherent to the design of DALI, but highlight them to
illustrate the differences between inference and training of DNNs.

\begin{figure}[t!]
  \includegraphics[width=\columnwidth]{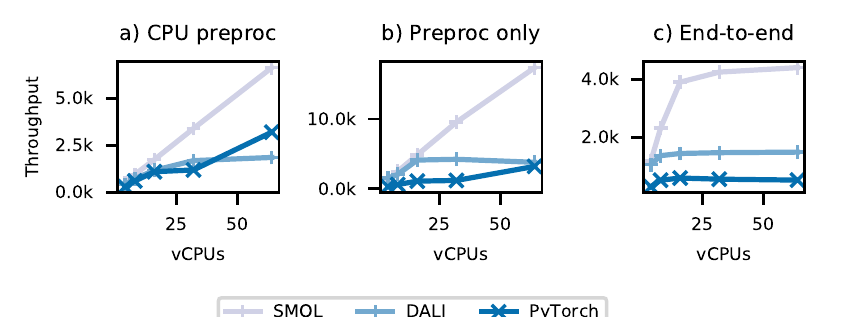}
  \caption{DALI, PyTorch, and \sn for a) CPU preprocessing (\sn
  optimizations off), b) optimized preprocessing, and c) end-to-end DNN
  inference. \sn outperforms both PyTorch and DALI in all settings
  except low vCPU for optimized preprocessing.}
  \label{fig:dali}
\end{figure}

We first compare DALI's and PyTorch's CPU implementation of preprocessing to
\sn's CPU implementation of preprocessing with all optimizations turned off. As
shown in Figure~\ref{fig:dali}, \sn's CPU implementation outperforms both DALI's
and PyTorch's CPU implementation for any number of cores. Due to
differences in implementation, we were unable to determine the cause of the
difference. Nevertheless, we hypothesize that the primary differences against
DALI occur in memory allocation: \sn reuses buffers while DALI must allocate new
buffers (as is required for integration with DNN training libraries).
PyTorch's inefficiency at 32 cores is likely due to lack of NUMA
awareness.

We next compare DALI's split CPU/GPU preprocessing implementation with \sn's
optimized implementation. We note that DALI does not optimize across
preprocessing and downstream tasks, so it uses a fixed pipeline regardless of
the number of cores. As shown in Figure~\ref{fig:dali}, \sn outperforms DALI from 8
vCPU cores. We hypothesize that DALI outperforms \sn at 4 vCPU cores due to
differences in resize implementations. We further hypothesize that \sn outperforms
DALI at higher vCPU cores due to GPU access contention in DALI.

Finally, we compare end-to-end inference of DALI, PyTorch, and \sn. As
noted, DALI does not officially support integration with TensorRT. Thus,
integrating DALI and TensorRT requires additional memory copies compared to \sn.
As shown in Figure~\ref{fig:dali}, \sn outperforms DALI in all settings, due to
the additionally memory copies. Both DALI and \sn outperform PyTorch as
PyTorch does not use an optimized DNN inference compiler.

\end{appendix}

\end{document}